# Optimal optical polarization of nitrogen-vacancy center with arbitrary waveform pulse

JIXING ZHANG,[1] TIANZHENG LIU,[1] LIXIA XU,[1] GUODONG BIAN,[1] PENGCHENG FAN,[1] MINGXIN LI[1], NING ZHANG[5] AND HENG YUAN,[1,2,3,4*]

[1]*School of Instrumentation and Optoelectronic Engineering, Beihang University, Beijing 100191, China;*
[2]*Research Institute of Frontier Science, Beihang University, Beijing 100191, China;*
[3]*Beijing Advanced Innovation Center for Big Data-Based Precision Medicine, Beihang University, Beijing 100191, China;*
[4]*Beijing Academy of Quantum Information Sciences, Beijing 100193, China;*
[5]*Zhejiang Lab, Hangzhou, 310000, China;*
*\*hengyuan@buaa.edu.cn,*

**Abstract:** The current work proposes a method for pulsed-light polarization of nitrogen-vacancy (NV) center electron spin. To evaluate the influence of pulsed spin polarization, we establish a polarization evaluation index based on polarizability and polarization time. Master equation model are utilized to theoretically calculate the spin polarization dynamics under light excitation and the optimal polarization conditions for the conventional methods are obtained. A novel pulsed-light polarization method is proposed by changing the optical pumping rate in the master equation from a fixed value to a time variable and an optimal waveform for proposed method is demonstrated through the variational method, which can simultaneously achieve high polarizability and requires a short polarization time. Hence, the polarization evaluation index is improved by ~10%. Moreover, the proposed method is verified by a pulsed-laser experimental system based on an arbitrary waveform generator. The current report shall expand the application horizon of NV center based quantum sensing.

## 1. Introduction

The nitrogen-vacancy (NV) center in diamond is a solid spin material, which has garnered significant research interest in recent years due to optical polarization and readout of the spin state. NV center exhibits promise in a wide variety of applications, such as magnetic field measurements [1][2], temperature measurements [3], rotation sensing [4][5], nanoscale nuclear magnetic resonance [6] and biological research [7]. The NV center in diamond can be used for quantum sensing applications where traditional sensors cannot work effectively, such as extremely high pressure [8], cells [9][10], distributed quantum measurements in fibers [11], and nanoscale magnetic imaging via an atomic force microscope (AFM) [12]. The quantum sensing using NV centers can be realized using a continuous optical detected magnetic resonance (ODMR) scheme [13], which is similar to the traditional optical pump magnetometers. Alternatively, quantum sensing can also be achieved via pulsed schemes, where light does not interfere with spin manipulation and renders broad application prospects.

In general, the pulsed NV center measurements require three steps [14]: initialization of polarizing NV center, spin-state manipulation and spin-state readout. The techniques for improving manipulation and readout have been extensively studied [15]. For instance, the spin-state manipulation techniques include microwave techniques [16][17] and spin dynamic decoupling sequence techniques [18][19], whereas the readout techniques include spin readout algorithms [20][21] and fluorescence collection techniques [22]. However, the improvement of polarization has rarely been studied.

As polarization is a prerequisite for spin-state manipulation and of great significance for enhancing the sensitivity of NV-based quantum sensing, we aimed to improve the polarization efficiency of NV center pulse optical polarization. The conventional NV center polarization utilizes a optical pulse to excite the NV center and spin-related relaxation of the excited states and singlet to realize the NV center spin polarization to $m_s = 0$ [23]. There are several theoretical and experimental studies on the physical principles of optical polarization, including theoretical analysis of the light-induced spin polarization [24][25], intersystem crossing in NV center and its effect on polarization [26][27], relaxations from singlet to ground state and their effect on polarization [28], and analysis of polarization characteristics using femtosecond pulses[29]. The polarizability can be defined as the proportion of a certain spin substate ($m_s = 0$) in the ground state. In general, high polarizability and short polarization time cost are conducive to the improved performance.

Therefore, we have established an evaluation index, combining the polarizability and polarization time, to evaluate the polarization efficiency. Through theoretical calculations, we have demonstrated that the polarizability and polarization speed contradicts each other for the conventional methods. For instance, the higher polarizability requires longer polarization time, which implies that the evaluation index of the corresponding polarization efficiency is not good enough. Therefore, based on the characteristics of optical polarization dynamics and variational method, we have designed a pulsed optical polarization scheme with time-variant pumping rate during the pulse process. We have demonstrated the optimal optical pumping rate function and improved the optimal evaluation index by 10% compared with the conventional methods.

## 2. Theoretical model

The room-temperature energy level structure and optical dynamics of the NV center have been theoretically and experimentally evaluated in detail [30], as shown in Fig. 1(a). The ground state is spin-triplet with a zero-field spilt ($Z_{gs} = 2.87$ GHz) between substate-1 ($m_s = 0$) and substate-2 ($m_s = \pm 1$). The excited state is also a spin-triplet with substate-3 ($m_s = 0$) and substate-4 ($m_s = \pm 1$). The emission relaxation from the excited state to the ground state is spin-independent, and its rate is represented by $k$. This process produces 600-800 nm fluorescence. Typically, the ground state is spin-independently pumped to the excited state non-resonantly by a 532 nm laser. The pumping rate is represented by $\Gamma$, which is proportional to the optical power density of the excitation light. The singlets between the excited state and ground state are merged into one singlet-5. The $k_{35}$ and $k_{45}$ relaxations from the excited state to singlet and $k_{51}$ and $k_{52}$ relaxations from singlet to ground state are spin-dependent. These two spin-dependent decays are responsible for the optical polarization spin of the NV center. The longitudinal relaxation time ($T_1$) at room temperature is ~1 ms.

The dephasing time ($T_2^*$) is not considered in optical dynamics. According to the given model, we can use the master equation to build a mathematical model as follows:

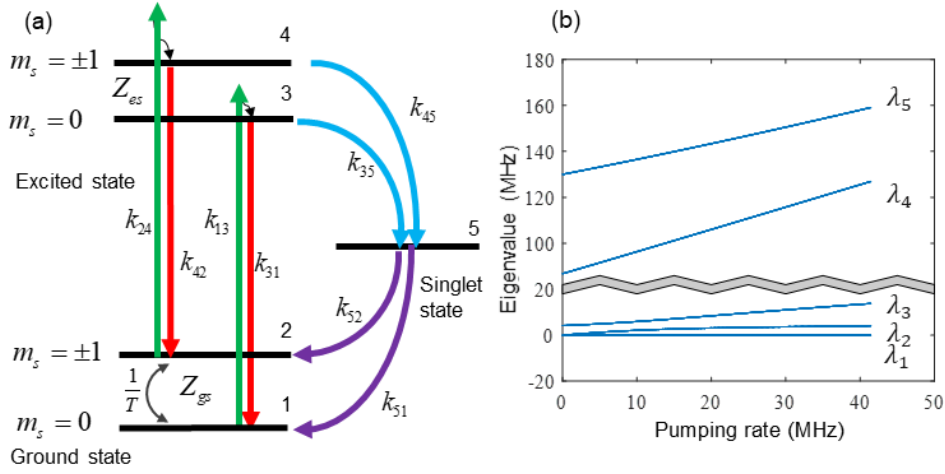

Figure 1 (a) Energy level structure of the NV center. (b) the change in five eigenvalues with respect to the optical pumping rate ($\Gamma$).

$$\frac{d\rho_1}{dt} = -\Gamma\rho_1 + k\rho_3 + k_{51}\rho_5 - \frac{1}{T_1}(\rho_1 - \rho_2)$$

$$\frac{d\rho_2}{dt} = -\Gamma\rho_2 + k\rho_4 + k_{52}\rho_5 + \frac{1}{T_1}(\rho_1 - \rho_2)$$

$$\frac{d\rho_3}{dt} = \Gamma\rho_1 - k\rho_3 - k_{35}\rho_3 \qquad (1)$$

$$\frac{d\rho_4}{dt} = \Gamma\rho_2 - k\rho_4 - k_{45}\rho_4$$

$$\frac{d\rho_5}{dt} = k_{35}\rho_3 + k_{45}\rho_4 - k_{51}\rho_5 - k_{52}\rho_5$$

Herein, $\rho_i$ denotes the $i^{\text{th}}$ state's probability distribution, which implies that $\sum \rho_i = 1$. Eq. (1) can also be rewritten in the matrix form:

$$\frac{d\vec{\rho}}{dt} = (A_0 + B\Gamma)\vec{\rho} \qquad (2)$$

$A_0$ and $B$ can be given as:

$$A_0 = \begin{bmatrix} -1/T_1 & 1/T_1 & k & 0 & k_{51} \\ 1/T_1 & -1/T_1 & 0 & k & k_{52} \\ 0 & 0 & -k-k_{35} & 0 & 0 \\ 0 & 0 & 0 & -k-k_{45} & 0 \\ 0 & 0 & k_{35} & k_{45} & -k_{51}-k_{52} \end{bmatrix}, \quad B = \begin{bmatrix} -1 & 0 & 0 & 0 & 0 \\ 0 & -1 & 0 & 0 & 0 \\ 1 & 0 & 0 & 0 & 0 \\ 0 & 1 & 0 & 0 & 0 \\ 0 & 0 & 0 & 0 & 0 \end{bmatrix} \qquad (3)$$

## 3. Polarization process, steady polarizability and polarization evaluation index

Optical dynamics, determined by Eq. (2), are mathematically solvable [20]. For the conventional scheme, where Γ remains constant, the optical dynamics can be given as:

$$\vec{\rho}(t) = e^{(A_0 + B\Gamma)t} \vec{\rho}(0) \quad (4)$$

Where,

$$e^{(A_0 + B\Gamma)t} = \begin{bmatrix} e^{\lambda_1 t}\vec{u}_1 & \cdots & e^{\lambda_5 t}\vec{u}_5 \end{bmatrix} \begin{bmatrix} \vec{u}_1 & \cdots & \vec{u}_5 \end{bmatrix}^{-1} \quad (5)$$

$\lambda_1, \cdots, \lambda_5$ denote the five different eigenvalues of $(A_0 + B\Gamma)$, where the corresponding eigenvectors can be given as: $\vec{u}_i = [u_i^1, u_i^2, u_i^3, u_i^4, u_i^5]^T$. For the pulse scheme, the initial distribution of the excited states and singlet state is 0. Thus, the initial $\vec{\rho}(0)$ is only determined by polarizability $P_0$ at the ground state, which can be defined as the population of $\rho_1$. For the initial state before polarization, $P_0 = 0.5$. By setting $w_j^i$ the $ij^{th}$ element of $[\vec{u}_1, \vec{u}_2, \vec{u}_3, \vec{u}_4, \vec{u}_5]^{-1}$ [21], the polarizability can be given as:

$$P(t) = \frac{1}{2} \sum_{i=1}^{i=5} e^{\lambda_i t} u_i^1 (w_1^i + w_2^i) \quad (6)$$

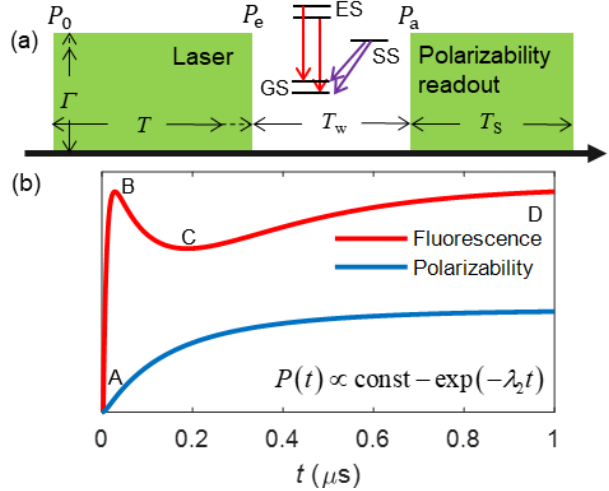

Figure 2 (a) Polarization process of the conventional method, where $P_0$ refers to the initial polarizability and the polarization process consists of a light part and a dark part. The time of the light part is denoted by $T$ and the polarizability at the end of light part is defined as $P_e$. The time of the dark part is represented by $T_w$ and both excited and singlet states are relaxed to the ground state during the dark part. At the end of dark part, the polarizability ($P_a$) refers to the final polarizability of the polarization process. The second light pulse is for spin readout pulse and the time is defined as $T_S$. (b) The simulated fluorescence dynamics and polarization dynamics at $\Gamma = k$. The fluorescence is influenced by different exponential processes, where the AB segment is mainly determined by the exponential process of $\lambda_4$ and $\lambda_5$, BC segment is dictated by the exponential process of $\lambda_3$, and CD segment is defined by the exponential process of $\lambda_2$. After D, the steady-state process is rendered by $\lambda_1 = 0$.

As shown in Fig. 1(b), $\lambda_3$, $\lambda_4$, and $\lambda_5$ are significantly higher than $\lambda_2$ and $\lambda_1 = 0$. Thus, the dynamics of polarizability longer than 100 ns is mainly affected by $\lambda_2$, thus, $\lambda_2$ is defined as polarization speed as shown in Fig. 2(b). In the actual spin operation, the spin manipulation occurs in the ground state. However, when the laser pulse ends, the non-ground states are still relaxing to the ground states. Therefore, the actual polarizability should be defined as the $\rho_i$ after a dark time that allows complete relaxation of all non-ground states into the ground states, as shown in Fig. 2(a). According to the value of $k$, $k_{51}$ and $k_{52}$, the dark time is set to $T_w = 2$ μs. Therefore, the actual polarizability should be,

$$P_a = \rho_1 + \rho_3 \frac{k_{31}}{k_{35}+k_{31}} + \left(\rho_3 \frac{k_{31}}{k_{35}+k_{31}} + \rho_4 \frac{k_{45}}{k_{45}+k_{42}} + \rho_5\right)\frac{k_{51}}{k_{51}+k_{52}} \quad (7)$$

By setting $k_{35}/k_{45} = \phi$, $k_{51}/k_{52} = \varphi$, and $T_1$ to be infinity, we can calculate the polarizability limit as:

$$P_e^{T_1 \to \infty} = \frac{1}{(1+\phi\varphi)+\dfrac{\varphi}{k/k_{35}+1}} \quad (8)$$

The polarization speed increases and $P_e^{T_1 \to \infty}$ decreases with increasing $\Gamma$. The maximum value of $P_e^{T_1 \to \infty}$ occurs at $\Gamma = 0$, as shown in Fig. 3 (a). Hence, the maximum polarizability limit is achieved when $\Gamma$ is infinitely small and polarization time is infinitely long. At $T_1 = 1$ ms, the actual polarizability (solid blue curve) cannot reach the polarizability limit at low $\Gamma$ values because $1/T_1 \leq \Gamma$ and the effect of $T_1$ cannot be ignored.

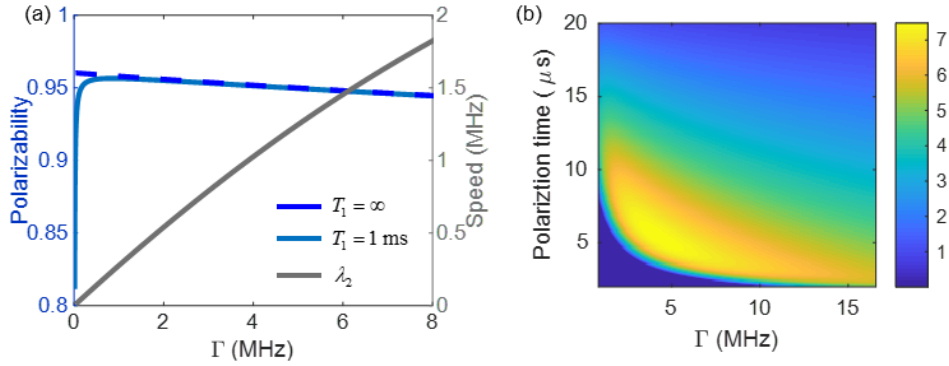

Figure 3 (a) Polarizability and polarization speed with respect to $\Gamma$. The dotted and solid blue-colored curves represent the polarizability where $T_1$ decoherence is infinitely long and 1 ms, respectively. (b) The simulated optimal polarization is achieved at the polarization time of 5.94 μs, polarizability of 0.9511, pumping rate of 3.64 MHz and polarization evaluation index of 7.4780.

The established model can analyze the optimal polarization conditions. First, we have set up an evaluation index for the polarization results. The optimization parameters are $T$ and $\Gamma$. According to the above calculations, the polarization speed $\lambda_2$ is related to $\Gamma$, however, the polarization time $T$ can be selected independently. Hence, $T$ and $\Gamma$ are not particularly related. In order to evaluate the polarization efficiency, we have established an evaluation parameter based on $T$ and $\Gamma$. The parameter renders optimal polarization efficiency when corresponding magnetic field measurement sensitivity is maximum. As the total time is $T+T_W+T_S$, the theoretical sensitivity limit of the magnetic field measurement can be given as:

$$\eta = \frac{\sqrt{T+T_W+T_S}}{C(P_a)RT_2^*} \text{ [14]}.$$

where $R$ denotes the light collection efficiency, $C$ is the spin readout contrast, which is proportional to Pa. Therefore, the evaluation index can be given as:

$$S = \frac{P_a - P_{\min}}{\sqrt{T+T_W+T_S}} \tag{9}$$

The optimal polarization maximizes $S$ by solving the following optimization problem:

$$\text{Maximize: } S = \frac{P_a(\Gamma, T) - P_{\min}}{\sqrt{T+T_W+T_S}}$$
$$\text{Subject to: } \begin{array}{l} 0 < T < T_{\max} \\ P_a > P_{\min} \end{array} \tag{10}$$

where, $P_{\min}$ represents the lowest allowable polarizability and $T_{\max}$ represents the maximum allowable polarization time. The simulated results of optimal conditions for the conventional method are shown in Fig. 3(b). The optimal polarization occurs under the polarization time of 5.9 μs and a pumping rate of 3.64 MHz. Fig. 3(b) shows that too long $T$ and too high $\Gamma$ reduced the polarization evaluation index because the higher $\Gamma$ increased the polarization speed and reduced $T$, however, it also compromised the polarizability.

## 4. Improvement methods and optimal results

The conventional method is based on a square-wave laser pulse. Since $\Gamma$ is constant, the polarization speed ($\lambda_2$) remains unchanged. Hence, the polarization time and polarizability are contradictory variables. The high polarization speeds require high pumping rate, whereas the high pumping rate compromises the final polarizability. To solve the dilemma, the proposed improvement method transformed the constant $\Gamma$ into time-dependent $\Gamma(t)$. Therefore, we need to find an optimal $\Gamma(t)$ function to maximize the polarization evaluation index. The maxima of a variable can be obtained by finding the position where the derivative is 0. Similarly, the maxima of a function can be obtained by finding the position where the variation is 0, known as calculus of variations. At a certain $T$, the optimal $S$ is obtained by finding optimal $P_a(T)$. The corresponding $L$ function satisfies the given relationship:

$$P_a(t_0) = \int_0^T L(\Gamma, \Gamma', t) dt \tag{11}$$

Obviously, $L$ is equal to the derivative of $P_a$ and, according to Eq. (7), $L$ can be defined as:

$$L = P_a'(\Gamma, \Gamma', t) = D \cdot \vec{\rho}'(t) \tag{12}$$

where

$$D = \left(1, 0, \frac{k_{31}}{k_{35}+k_{31}}\left(1+\frac{k_{51}}{k_{51}+k_{52}}\right), \frac{k_{45}k_{51}}{(k_{45}+k_{42})(k_{51}+k_{52})}, \frac{k_{51}}{k_{51}+k_{52}}\right) \tag{13}$$

The aimed $P_a'$ satisfies the Euler–Lagrange equation:

$$\frac{\partial P_a{'}(\Gamma,\Gamma',t)}{\partial \Gamma} - \frac{d}{dt}\left(\frac{\partial P_a{'}(\Gamma,\Gamma',t)}{\partial \Gamma'}\right) = 0 \tag{14}$$

Eq. (14) can be solved by Eq. (4), as given below:

$$P_\alpha{'} = D(A_0 + B\Gamma)\exp((A_0 + B\Gamma)t)\vec{\rho}_0 \tag{15}$$

The optimal $\Gamma(t)$, obtained from Eq. (14), is shown as the dark-green curve in Fig. 4(a), which is a concave function and exhibits a strong pumping rate at begin and decreases with time. The physical mechanism for the variational result is that the high $\Gamma$ at begin increases the polarization speed, which leads to high polarizability in a short time, and low $\Gamma$ at end further increases $P_a$. Therefore, the concave function is more suitable than the convex function. However, the given curve does not represent the final optimal waveform because the variational method only considers increasing $P_a$ and ignores the reducing polarization time. Moreover, the results of variational method do not depend on the boundary conditions of the polarization time, hence, the optimal polarization time cannot be achieved by the given variational method.

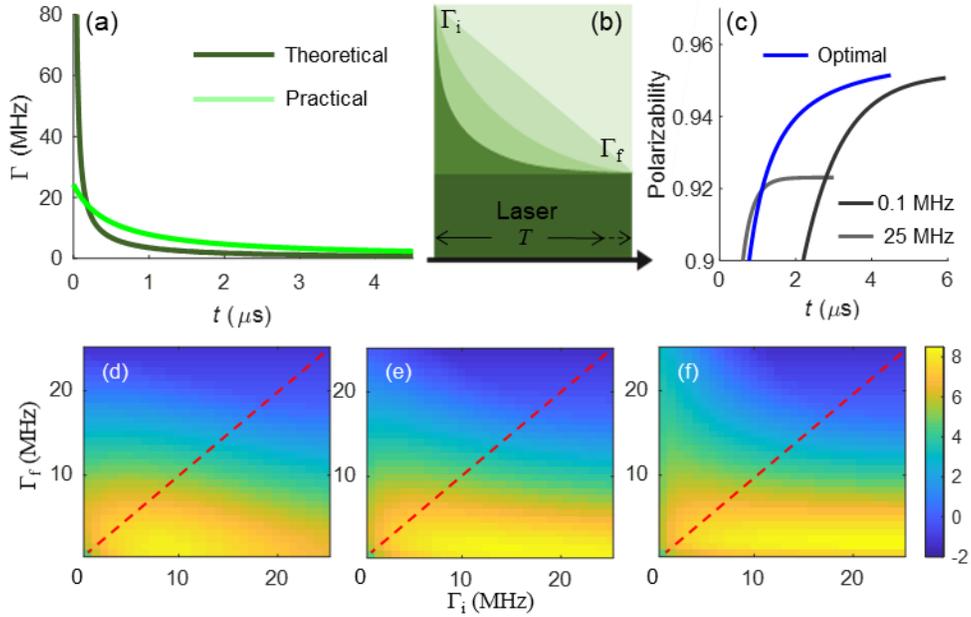

Figure 4 (a) The theoretical optimal optical pumping rate waveform obtained by the variational method (dark-green curve) and the actual optimal optical pumping rate waveform based on three waveform simulations (light-green curve). (b) Three waveforms were designed according to the idea of variational method. The initial optical pumping rate is $\Gamma_i$, the end optical pumping rate is $\Gamma_f$, and the polarization time is T. (c) The polarizability of different methods increased with time, where the blue-colored curve corresponds to the $\Gamma(t)$ of the light-green curve in (a), and the dark-gray and light-gray curves correspond to the conventional method with $\Gamma = 25$ MHz and $\Gamma = 0.1$ MHz, respectively. Consistent with the theoretical analysis, the optimal method rendered fast polarization speed and high polarization rate. (d) The optimal conditions for triangular waves: optimal polarization evaluation index: 8.1358, $\Gamma_i = 8.1$ MHz, $\Gamma_f = 1.54$ MHz, polarization time: 5.1 μs, and polarizability: 0.951. (e) The optimal conditions for exponential waves: optimal polarization evaluation index: 8.3457, $\Gamma_i = 13.8$ MHz, $\Gamma_f = 1.54$ MHz, polarization time: 4.8 μs, and polarizability: 0.9518. (f) The optimal conditions for inverse proportional wave: optimal polarization evaluation index: 8.4289, $\Gamma_i$ = 24.3 MHz, $\Gamma_f$ = 2.35 MHz, polarization time: 4.5 μs, and polarizability: 0.9515.

Therefore, we have further analyzed the optimal polarization method based on the results of the variational method by utilizing three different monotonically decreasing concave functions to approximate the optimal function, i.e., triangular, exponential and inverse proportional functions (Fig. 4b). Unlike conventional square waves, the proposed method cannot be directly solved because $\Gamma(t)$ is constantly changing. The shape of $\Gamma(t)$ waveform can be fully characterized by the starting pumping rate $\Gamma_i$, end pumping rate $\Gamma_f$ and $T$. The triangular waveform function can be given as:

$$\Gamma(t) = \frac{\Gamma_f - \Gamma_i}{T} t + \Gamma_i \tag{16}$$

The exponential waveform function can be given as

$$\Gamma(t) = \Gamma_i \exp\left( \log\left(\frac{\Gamma_f}{\Gamma_i}\right) \frac{t}{T} \right) \tag{17}$$

The exponential waveform function can be given as:

$$\Gamma(t) = \frac{\Gamma_i T}{T + \left(\Gamma_i / \Gamma_f - 1\right) t} \tag{18}$$

The simulation results are presented in Fig. 4(d-f). The red-colored dotted line is obtained when $\Gamma_i = \Gamma_f$, which corresponds to the conventional fixed $\Gamma$ method. The optimal polarization evaluation index appeared below the red-colored dotted line. Consistently, the optimal polarization position occurs when the initial pumping rate is high and the end pumping rate is low. Among the three waveform functions, the inverse proportional function exhibited the optimal results, as shown by the light-green curve in Fig. 4(a). Compared with the traditional method, it achieves an increase in sensitivity and a significant reduction in polarization time, as shown in Fig. 4(c).

## 5. Experiment verification

The setup of the verification experiment should realize the generation of arbitrary waveform laser pulses and independence between polarization and detection pulses. The as-designed experimental setup is shown in Fig. 5(a). To achieve ideal polarization, a time variant waveform pulse was generated by an arbitrary waveform generator (AWG7122b) and a pulse generator (Spincore). The corresponding laser pulse was achieved by an acousto-optic modulator (AOM) (Gooch&Housego 3200-1911), with an analog driver (1200AF-AIFO-1.0), and an AOM with a digital driver (1200AF-DIFO-1.0). The power of the excitation laser source (Oxxius LCX-532) and detection laser source (MLL-U-532-500mW) reached 500 mW. The laser power can be reduced by a combination of λ/2 wave plate mounted on a motorized precision rotation mount (Thorlab PRM1/MZ8) and polarization beam splitter (PBS). The objective lens was 80×/0.8 with a focal length of 0.17 mm. An avalanche photodiode (APD, Thorlab APD120A2/M) was used for fluorescence detection. Home-made Helmholtz coils produced a static magnetic field. The microwave source was SSG-6000RC and the microwave pulse was generated by a Spincore controlled microwave switch (ZASW-2-50DR+). The overall experimental setup is shown in Fig. 5(a) and (b). To verify the versatility of the improved method, we have utilized three NV center diamond samples from high to low concentrations, as shown in Fig. 5(c).

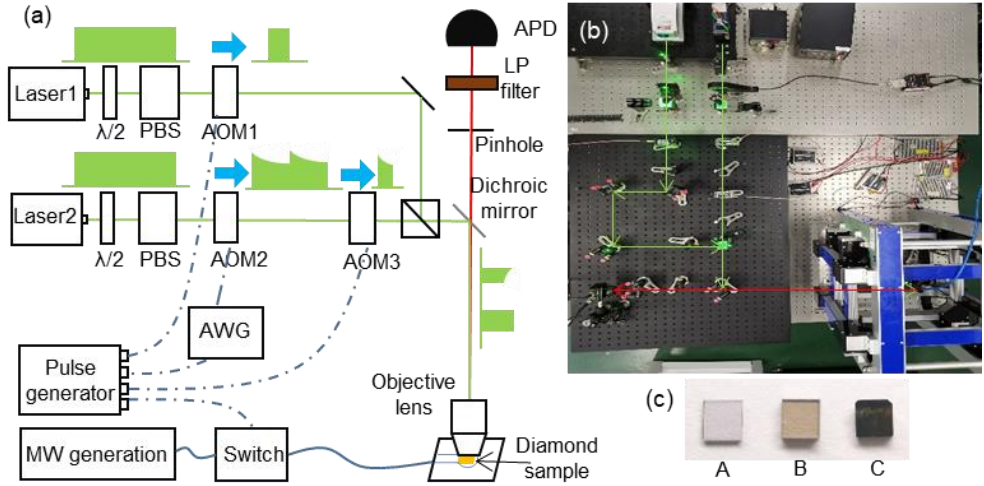

Figure 5 (a) Schematic diagram of the experimental setup and (b) digital photograph of the experimental apparatus. The detection laser changed from continuous light to pulsed-light through a digital AOM. The laser was first transformed into a periodic waveform through an analog AOM, controlled by the AWG, and transformed into the required pulsed waveform through the pulsed AOM. The two lasers were combined through the BS. Hence, the independence of the detection pulse can be ensured during the change of polarization pulse, ensuring the consistency of spin readout. An objective lens of 80x was employed to achieve high optical power density. The fluorescence light path was spatially filtered through a pinhole. (c) Digital photographs of the samples, where Sample-A was prepared by chemical vapor deposition (CVD) of type-Ia diamond (Element Six) with N impurity concentration of <1 ppm. The Sample-A was treated under $1\times10^{17}$ ea/cm$^2$ with 10 MeV electron irradiation and annealed at 800 °C for 2 h. The Sample-B was a high-pressure high-temperature (HPHT) type-Ib diamonds (Element Six) with N impurity concentration of <200 ppm. The Sample-B was treated under $5\times10^{17}$ ea/cm$^2$ with 10 MeV electron irradiation and annealed at 800 °C for 2 h. Sample-C is a CVD diamond with N impurity concentration of ~50 ppm. Sample-C was treated under $1\times10^{18}$ ea/cm$^2$ with 10 MeV electron irradiation and annealed at 800 °C for 3 h.

The maximum optical pumping rate $\Gamma$ corresponding to the maximum laser power of 500 mW was measured to be ~40 MHz by the measurement method from Ref [21]. The experimental sequence is shown in Fig. 6(a). A-B and C-D are to eliminate the influence of detection laser power fluctuations [20]. Since microwave flipping only influences the electron spin, (A-B) - (C-D) eliminates the influence of NV$^0$ dynamics [31]. For the given sequence, at $\Gamma_i=\Gamma_f$, the polarization waveform corresponds to the waveform of the conventional method. The experimental results of the polarization efficiency of the conventional method are shown in Fig. 6(e), which are consistent with the simulation results (Fig. 3b). The experimental results of the proposed polarization evaluation index in the inverse proportional function method are shown in Fig. 6(b), which are also consistent with the simulation results (Fig. 4f). We have independently analyzed the changes in polarizability and polarization evaluation index with respect to the polarization time under a fixed optimal $\Gamma_f$ and different $\Gamma_i$ conditions (Fig. 6c). As the initial pumping rate decreased, the polarization speed was significantly reduced, however, the final polarizability remained almost the same. The polarization evaluation index exhibited an optimal value with polarization time changing. These results are consistent with the previously reported theoretical calculations. The improvement in final optimal index of all three samples is shown in Fig. 6(f), which is quite close to the theoretical values.

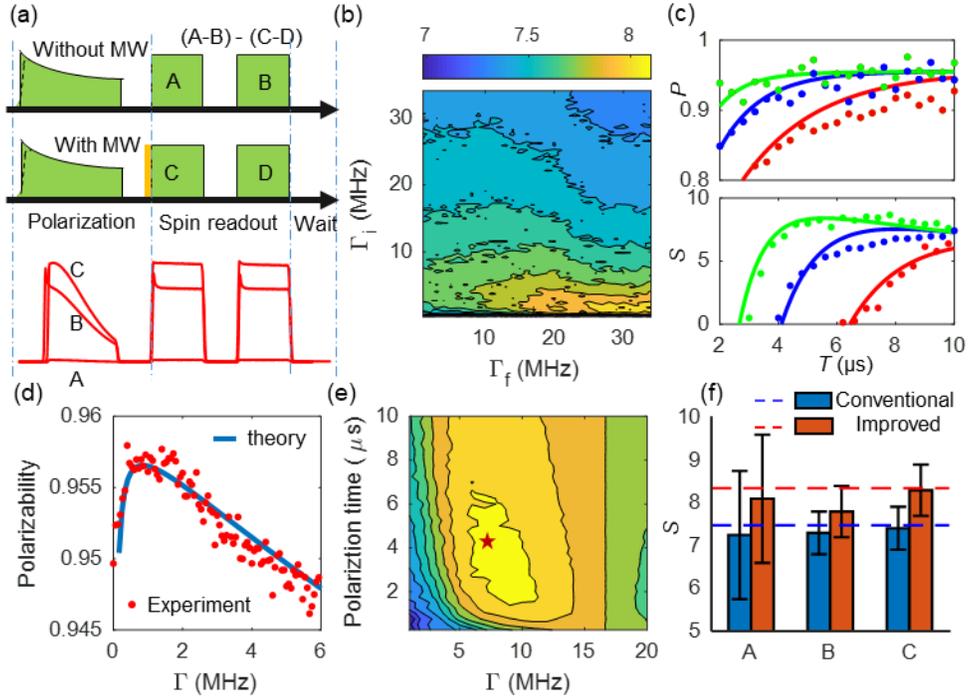

Figure 6 (a) A schematic illustration of the experimental sequence, where the green represents the laser pulse and orange represents the microwave pulse. After the polarization process, the detection part contains two laser pulses with an interval of 2 μs. The sequence above does not contain microwave pulses, whereas the sequence below adds a microwave π pulse between the polarization and detection pulse. The final polarizability result is (A-B)-(C-D). The lower part of (a) shows the corresponding fluorescence waveform, where different fluorescence values of three samples under the same optical pumping rate represent different NV center concentrations. Overall, the concentration ratio is around A:B:C =1:23:31. (b) The experimental polarization evaluation index of the inverse proportional method. The optimal polarization condition occurs at $\Gamma_f$ = 32 MHz and $\Gamma_f$ =1.5 MHz. (c) The change in polarizability (top) and polarization evaluation index (bottom) with the polarization time for the inverse proportional function method. The three curves were measured at $\Gamma_f$ = 2.35 MHz. The initial polarizability ($\Gamma_i$) was 3 MHz (red), 8 MHz (blue), and 25 MHz (green). (d) The experimental and theoretical results of the steady-state polarizability. The theoretical results are based on the data in Fig. 3(a), whereas the experimental results are obtained from Sample-C. (e) The experimental results of the conventional method of polarization evaluation index S with $\Gamma$ and $T$ for Sample-C. The optimal conditions are $T$ = 4.2 μs and $\Gamma$ = 7 MHz. (f) The comparison between the polarization of conventional (blue) and improved methods (red). Dot lines are theoretical results and histograms are experiments for different samples. Sample C showed a good match between theoretical and experimental results. Experimental polarization evaluation indexes for Sample B and A are worse than theoretical results for both methods, but the improved method are better than conventional method. Results for Sample A has large confidence intervals due to the low NV concentration and low fluorescence.

## 6. Conclusions

In summary, we have utilized the master equation model to analyze the polarization process of the NV center spin. It is demonstrated that the optical pumping rate plays a decisive role in the polarization process. Moreover, a polarization evaluation index consisting of optical pumping rate and polarization time has been proposed. Using the polarization evaluation index, we have theoretically calculated and experimentally measured the optimal polarization conditions for the conventional method. Based on the principle of variational method, we have proposed an improved polarization method using time-varying optical pumping rates. The theoretical and experimental results verify that the polarization evaluation index of the improved method has been increased by ~10%. Therefore, the proposed method can be widely used in high-performance diamond NV center quantum sensing.

**Funding**

This work was supported by the National Key R&D Program of China (2016YFB0501600); the Projects of National Natural Science Foundation of China under Grant No. 61773046 and 61721091; the Projects of Beijing Academy of Quantum Information Sciences under Grant No. Y18G33; Project funded by China Postdoctoral Science Foundation 2019M662121.